\documentclass{aa}
\usepackage{graphicx}
\usepackage{natbib}
\usepackage{bm,times,color}
\graphicspath{{./fig/}{./png/}}

\newcommand{\EQ}{\begin{equation}}
\newcommand{\EN}{\end{equation}}
\newcommand{\EQA}{\begin{eqnarray}}
\newcommand{\ENA}{\end{eqnarray}}

\newcommand{\Eq}[1]{Eq.~(\ref{#1})}

\newcommand{\Fig}[1]{Fig.~\ref{#1}}
\newcommand{\FFig}[1]{Figure~\ref{#1}}
\newcommand{\Figs}[2]{Figs.~\ref{#1} and \ref{#2}}
\newcommand{\Figss}[2]{Figs.~\ref{#1}--\ref{#2}}

\newcommand{\bra}[1]{\langle #1\rangle}

\newcommand{\meanrho}{\overline{\rho}}

{}
{}
{}
\newcommand{\meanemf}{\overline{\cal E} {}}

\newcommand{\meanSSSS}{\overline{\mbox{\boldmath ${\mathsf S}$}} {}}

\newcommand{\meanSSS}{\overline{\mathsf{S}}}
{}
{}
\newcommand{\meanEMF}{\overline{\mbox{\boldmath ${\cal E}$}}{}}{}
{}
{}
{}
{}
{}
\newcommand{\meanAA}{\overline{\mbox{\boldmath $A$}}{}}{}
\newcommand{\meanBB}{\overline{\mbox{\boldmath $B$}}{}}{}
{}
\newcommand{\meanFF}{\overline{\mbox{\boldmath $F$}}{}}{}
{}
{}
{}
{}
{}
{}
\newcommand{\meanJJ}{\overline{\mbox{\boldmath $J$}}{}}{}
{}
\newcommand{\meanUU}{\overline{\mbox{\boldmath $U$}}{}}{}

{}
{}
\newcommand{\meanQQ}{\overline{\mbox{\boldmath $Q$}}{}}{}

\newcommand{\meanB}{\overline{B}}

\newcommand{\meanU}{\overline{U}}

\newcommand{\meanJ}{\overline{J}}

{}

{}
{}

\newcommand{\xxx}{\hat{\mbox{\boldmath $x$}} {}}

\newcommand{\zzz}{\hat{\mbox{\boldmath $z$}} {}}

\newcommand{\bb}{\bm{b}}
\newcommand{\BB}{\bm{B}}

\newcommand{\uu}{\mbox{\boldmath $u$} {}}
\newcommand{\UU}{\mbox{\boldmath $U$} {}}

\newcommand{\JJ}{\mbox{\boldmath $J$} {}}

\newcommand{\AAA}{\mbox{\boldmath $A$} {}}
\newcommand{\aaaa}{\mbox{\boldmath $a$} {}}

\newcommand{\ff}{\mbox{\boldmath $f$} {}}

\newcommand{\FF}{\mbox{\boldmath $F$} {}}

\newcommand{\QQ}{\mbox{\boldmath $Q$} {}}
\newcommand{\grav}{\mbox{\boldmath $g$} {}}
\newcommand{\nab}{\mbox{\boldmath $\nabla$} {}}
\newcommand{\OO}{\bm{\Omega}}

\newcommand{\oo}{\mbox{\boldmath $\omega$} {}}

\newcommand{\SSSS}{\mbox{\boldmath ${\sf S}$} {}}

\newcommand{\DD}{{\rm D} {}}

\def\la{\mathrel{\mathchoice {\vcenter{\offinterlineskip\halign{\hfil
$\displaystyle##$\hfil\cr<\cr\sim\cr}}}
{\vcenter{\offinterlineskip\halign{\hfil$\textstyle##$\hfil\cr<\cr\sim\cr}}}
{\vcenter{\offinterlineskip\halign{\hfil$\scriptstyle##$\hfil\cr<\cr\sim\cr}}}
{\vcenter{\offinterlineskip\halign{\hfil$\scriptscriptstyle##$\hfil\cr<\cr\sim\cr}}}}}
\def\ga{\mathrel{\mathchoice {\vcenter{\offinterlineskip\halign{\hfil
$\displaystyle##$\hfil\cr>\cr\sim\cr}}}
{\vcenter{\offinterlineskip\halign{\hfil$\textstyle##$\hfil\cr>\cr\sim\cr}}}
{\vcenter{\offinterlineskip\halign{\hfil$\scriptstyle##$\hfil\cr>\cr\sim\cr}}}
{\vcenter{\offinterlineskip\halign{\hfil$\scriptscriptstyle##$\hfil\cr>\cr\sim\cr}}}}}
\def\Ma{\mbox{\rm Ma}}
\def\Co{\mbox{\rm Co}}

\def\Pm{\mbox{\rm Pr}_M}
\def\Rm{\mbox{\rm Re}_M}

\def\Rey{\mbox{\rm Re}}

\def\Co{\mbox{\rm Co}}
\def\Gr{\mbox{\rm Gr}}

\def\cs{c_{\rm s}}

\def\qpz{q_{\rm p0}}
\def\qp{q_{\rm p}}
\def\betap{\beta_{\rm p}}

\def\betastar{\beta_{\star}}
\def\Peff{{\cal P}_{\rm eff}}

\def\kf{k_{\rm f}}

\def\epsk{\epsilon_{\it k}}
\def\epsf{\epsilon_{\rm f}}
\def\epsfz{\epsilon_{\rm f0}}

\def\urms{u_{\rm rms}}

\def\nut{\nu_{\rm t}}

\def\nuT{\nu_{\rm T}}

\def\etat{\eta_{\rm t}}
\def\etatz{\eta_{\rm t0}}
\def\etatz{\eta_{\rm t0}}

\def\etaT{\eta_{\rm T}}

\def\Beq{B_{\rm eq}}
\def\Beqz{B_{\rm eq0}}

\def\half{{\textstyle{1\over2}}}

\def\onethird{{\textstyle{1\over3}}}

\newcommand{\kG}{\,{\rm kG}}

\newcommand{\s}{\,{\rm s}}
\newcommand{\h}{\,{\rm h}}
\newcommand{\days}{\,{\rm days}}

\newcommand{\Mm}{\,{\rm Mm}}

\newcommand{\yapj}[3]{ #1, {ApJ,} {#2}, #3}

\newcommand{\yapjl}[3]{ #1, {ApJ,} {#2}, #3}

\newcommand{\yan}[3]{ #1, {Astron.\ Nachr.,} {#2}, #3}

\newcommand{\yana}[3]{ #1, {A\&A,} {#2}, #3}

\newcommand{\ygafd}[3]{ #1, {Geophys.\ Astrophys.\ Fluid Dyn.,} {#2}, #3}

\newcommand{\ysovl}[3]{ #1, {Sov.\ Astron.\ Lett.,} {#2}, #3}
\newcommand{\yjetp}[3]{ #1, {Sov.\ Phys.\ JETP,} {#2}, #3}

\newcommand{\yprl}[3]{ #1, {Phys.\ Rev.\ Lett.,} {#2}, #3}

\newcommand{\ymn}[3]{ #1, {MNRAS,} {#2}, #3}
\newcommand{\ynat}[3]{ #1, {Nature,} {#2}, #3}

\newcommand{\ysph}[3]{ #1, {Solar Phys.,} {#2}, #3}

\newcommand{\ypre}[3]{ #1, {Phys.\ Rev.\ E,} {#2}, #3}

\newcommand{\yjour}[4]{ #1, {#2}, {#3}, #4}

\newcommand{\ybook}[3]{ #1, {#2} (#3)}
\newcommand{\yproc}[5]{ #1, in {#3}, ed.\ #4 (#5), #2}

\newcommand{\smn}[2]{ #1, {MNRAS}, submitted, arXiv:#2}
\titlerunning{}
\authorrunning{S. Jabbari et al.}
\title{Magnetic flux concentrations from dynamo-generated fields}
\author{S. Jabbari\inst{1,2}\and A. Brandenburg\inst{1,2}\and
I. R. Losada\inst{1,2}\and N. Kleeorin\inst{3,1,4}\and I. Rogachevskii\inst{3,1,4}
}
\institute{
Nordita, KTH Royal Institute of Technology and Stockholm University,
Roslagstullsbacken 23, 10691 Stockholm, Sweden
\and
Department of Astronomy, AlbaNova University Center,
Stockholm University, 10691 Stockholm, Sweden
\and
Department of Mechanical Engineering, Ben-Gurion University of the Negev,
POB 653, Beer-Sheva 84105, Israel
\and
Department of Radio Physics, N.~I.~Lobachevsky State University of
Nizhny Novgorod, Russia
}
\date{\today,~ $ $Revision: 1.148 $ $}
\begin{document}

\abstract{
The mean-field theory of magnetized stellar convection gives rise to two
distinct instabilities: the large-scale dynamo instability, operating
in the bulk of the convection zone and a negative effective magnetic
pressure instability (NEMPI) operating in the strongly stratified surface layers.
The latter might be important in connection with magnetic spot formation.
However, the growth rate of NEMPI
is suppressed with increasing rotation rates.
On the other hand, recent direct numerical simulations (DNS) have shown
a subsequent increase in the growth rate.
}{
We examine quantitatively whether this increase in the growth rate of NEMPI
can be explained by an $\alpha^2$ mean-field dynamo, and whether
both NEMPI and the dynamo instability can operate at the same time.
}{
We use both DNS and mean-field simulations (MFS) to solve the underlying
equations numerically either with or without an imposed horizontal field.
We use the test-field method to compute relevant dynamo coefficients.
}{
DNS show that magnetic flux concentrations are still possible up to
rotation rates above which the large-scale dynamo effect produces
mean magnetic fields.
The resulting DNS growth rates are quantitatively reproduced with MFS.
As expected for weak or vanishing rotation, the growth rate of NEMPI
increases with increasing gravity, but there is a correction term for
strong gravity and large turbulent magnetic diffusivity.
}{
Magnetic flux concentrations are still possible for rotation rates
above which dynamo action takes over.
For the solar rotation rate, the corresponding turbulent turnover time
is about 5 hours, with dynamo action commencing in the layers beneath.
}

\maketitle

\section{Introduction}

The appearance of surface magnetic fields in the Sun presents some
peculiar characteristics, such as being strongly concentrated into
discrete spots.
The origin and depth of such magnetic flux concentrations has long been 
the subject of considerable speculation.
A leading theory by \cite{Par55} interprets the emergence of such spots
as the result of magnetically buoyant flux tubes at a depth of some 20\Mm.
This magnetic field must be the result of a dynamo,
but magnetic buoyancy also leads to the buoyant rise
and subsequent loss of those magnetic structures.
It was therefore
thought that the dynamo should operate mainly at
or even below the bottom of the convection zone
where magnetic buoyancy could be stabilized by a
subadiabatic temperature gradient \citep{Par75}.
This led eventually to the idea that sunspots
might be a direct consequence of dynamo-generated
flux tubes that rise all the way from the bottom
of the convection zone to the surface
\citep[e.g.,][]{Caligari}. However,
\cite{Sch80,Sch83} emphasized early on
that such fields would easily lose their systematic east--west
orientation while ascending through the turbulent
convection zone. \cite{DSC93} estimated that a
magnetic field strength of about $100\kG$ would
be needed to preserve the overall
east--west orientation \citep{Hale19} and also
to produce the observed tilt angle of active
regions known as Joy's law.

A great deal of effort has gone into determining the conditions under
which magnetic flux ropes may or may not be able to rise buoyantly
across the convection zone.
\cite{Emonet} determined for the first time
the basic minimum twist thresholds for the survival of
twisted magnetic flux ropes during the rise.
Subsequent studies were based on different types of numerical simulations,
which tested the underlying hypotheses and looked for other effects, such as the
robustness against background convective motions
\citep{Jou} and magnetic flux erosion by reconnection
with the background dynamo field \citep{Pinto}.
These studies, as well as many others
\citep[see, e.g.,][and references therein]{Fan08,Fan09}
specifically looked at which flux-rope configurations
are able to reproduce the observed emergent polarity tilt angles (Joy's law).

The observed variation in the number of sunspots in time and latitude
is expected to be linked to some kind of large-scale dynamo, as 
was modeled by \cite{Leighton} and \cite{SK69} long ago.
This led \cite{Sch80} to propose a so-called flux-tube dynamo approach
that would couple the buoyant rise of thin flux tubes to their regeneration.
However, even today the connection between dynamos and flux tubes is done
by hand \citep[see, e.g.,][]{CCJ07,MD14}, which means that an ad hoc
procedure is invoked to link flux tube emergence to a mean-field dynamo.
Of course, such tubes, or at least bipolar regions, should ultimately emerge from
a sufficiently well-resolved and realistic simulation of solar convection.
While global convective dynamo simulations of \cite{NBBMT11,NBBMT13,NBBMT14}
show magnetically buoyant magnetic flux tubes of $\approx40\kG$ field
strength, they do not yet address bipolar region formation.
Indeed, solar surface simulations of \cite{CRTS10} and \cite{RC14}
demonstrate that bipolar spots do form once a magnetic flux tube of
$10\kG$ field strength is injected at the bottom of their
local domain ($7.5\Mm$ below the surface).
On the other hand, the deep solar simulations of \cite{SN12} develop
a bipolar active region with just $1\kG$ magnetic field injected at the
bottom of their domain.
While these simulations taken together outline what might occur in
the Sun, they do not necessarily support the description of spots as a
direct result of thin flux tubes piercing the surface \citep[e.g.][]{Caligari}.

A completely different suggestion is that sunspots develop locally at the
solar surface, and that their east--west orientation would reflect the
local orientation of the mean magnetic field close to the surface.
The tilt angle would then be determined by latitudinal shear producing
the observed orientation of the meridional component of the magnetic
field \citep{B05}.
One of the possible mechanisms of local spot formation is the negative effective
magnetic pressure instability \citep[NEMPI; see][]{KRR89,KRR90,KR94,KMR96,RK07}.
Another potential mechanism of flux concentration is related to
a thermo-magnetic instability
in turbulence with radiative boundaries caused by
the suppression of turbulent heat flux through the large-scale
magnetic field \citep{KM00}.
The second instability has so far only been found
in mean-field simulations (MFS), but not
in direct numerical simulations (DNS)
nor in large-eddy simulations (LES).
By contrast, NEMPI has recently been found in DNS \citep{BKKMR11}
and LES \citep{Jab2} of strongly stratified fully developed turbulence.

As demonstrated in earlier work \citep{BKR13,Jab2},
NEMPI can lead to the formation of
equipartition-strength magnetic spots, which are
reminiscent of sunspots.
Even bipolar spots can form in the presence of a horizontal
magnetic field near the surface \citep[see][]{WLBKR13}.
For this idea to be viable, NEMPI and the dynamo instability
would need to operate in reasonable proximity to each other,
so that the dynamo can supply the magnetic field that would be
concentrated into spots, as was recently demonstrated by \cite{MBKR14}.
In studying this process in detail,
we have a chance of detecting new joint effects
resulting from the two instabilities,
which is one of the goals of the present paper.
However, these two instabilities may also compete against each other,
as was already noted by \cite{Los2}.
The large-scale dynamo effect relies on the combined presence of rotation
and stratification, while NEMPI requires
stratification and large enough scale separation.
On the other hand, even a moderate amount of rotation suppresses NEMPI.
In fact, \cite{Los1} found significant suppression of NEMPI
when the Coriolis number $\Co=2\Omega\tau$ is larger than about 0.03.
Here, $\Omega$ is the angular velocity and $\tau$ the turnover time of
the turbulence, which is related to the rms velocity $\urms$ and the
wavenumber $\kf$ of the energy-carrying eddies via $\tau=(\urms\kf)^{-1}$.
For the solar convection zone, the Coriolis number,
\EQ
\Co=2\Omega/\urms\kf,
\EN
varies from $2\times10^{-3}$ (at the surface using $\tau=5\min$)
to $5$ (at the bottom of the convection zone using $\tau=10\days$).
The value $\Co=0.03$
corresponds to a turnover time as short as two hours,
which is the case at a depth of $\approx10\Mm$.

The strength of stratification, on the other hand,
is quantified by the nondimensional parameter
\EQ
\Gr=g/\cs^2\kf \equiv (\kf H_\rho)^{-1},
\EN
where $H_\rho=\cs^2/g$ is the density scale height,
$\cs$ is the sound speed, and $g$ is the gravitational acceleration.
In the cases considered by \cite{Los1,Los2}, the stratification parameter
was $\Gr=0.03$, which is rather small compared with the estimated
solar value of $\Gr=0.16$ \citep[see the conclusions of][]{Los2}.
One can expect that larger values of $\Gr$ would result in correspondingly
larger values of the maximum permissible value of $\Co$, for which NEMPI
is still excited, but this has not yet been investigated in detail.

The goal of the present paper is to study rotating stratified
hydromagnetic turbulence in a parameter regime that we
expect to be at the verge between NEMPI and dynamo instabilities.
We do this by performing DNS and MFS.
In MFS, the study of combined NEMPI and dynamo instability requires
suitable parameterizations of the negative effective magnetic pressure
and $\alpha$ effects using suitable turbulent transport coefficients.

\section{DNS study}

We begin by reproducing some of the DNS results of \cite{Los2},
who found the suppression of the growth rate of NEMPI with increasing
values of $\Co$ and a subsequent enhancement
at larger values, which they
interpreted as being the result of dynamo action in the presence
of an externally applied magnetic field.
We also use DNS to determine independently the expected efficiency of
the dynamo by estimating the $\alpha$ effect from kinetic helicity
measurements and by computing both $\alpha$ effect and turbulent
diffusivity directly using the test-field method (TFM).

\subsection{Basic equations}

In DNS of an isothermally stratified layer \citep{Los2},
we solve the equations for the velocity $\UU$,
the magnetic vector potential $\AAA$, and the density $\rho$
in the presence of rotation $\Omega$,
\begin{eqnarray}
{\DD\UU\over\DD t}&=&{1\over\rho}\JJ\times\BB
-2\OO\times\UU-\nu\QQ+\FF+\ff,\\
{\partial\AAA\over\partial t}&=&\UU\times\BB-\eta\JJ,\\
{\partial\rho\over\partial t}&=&-\nab\cdot\rho\UU,
\end{eqnarray}
where $\DD/\DD t=\partial/\partial t+\UU\cdot\nab$ is the advective derivative,
$\OO=\Omega\zzz$ is the angular velocity,
\EQ
\FF=\grav-\cs^2\nab\ln\rho
\EN
determines the hydrostatic force balance,
$\nu$ is the kinematic viscosity, $\eta$ is the
magnetic diffusivity due to Spitzer conductivity of the plasma,
\EQA
-\QQ&=&\nabla^2\UU +\nab\nab\cdot\UU/3+2\SSSS\nab\ln\rho,\\
-\JJ&=&\nabla^2\AAA-\nab\nab\cdot\AAA,
\ENA
are the modified vorticity and the current density, respectively,
where the vacuum permeability $\mu_0$ has been set to unity,
\EQ
\BB=\BB_0+\nab\times\AAA
\EN
is the total magnetic field,  $\BB_0=(0,B_0,0)$ is the imposed uniform field, and
\EQ
{\sf S}_{ij}
=\half(\partial_j U_i+\partial_i U_j)-\onethird\delta_{ij}\nab\cdot\UU
\EN
is the traceless rate-of-strain tensor.
The forcing function $\ff$ consists of random, white-in-time,
plane, nonpolarized waves with a certain average wavenumber $\kf$.
The turbulent rms velocity is approximately
independent of $z$ with $\urms=\bra{\uu^2}^{1/2}\approx0.1\,\cs$.
The gravitational acceleration $\grav=(0,0,-g)$ is chosen such that
$k_1 H_\rho=1$, so the density contrast between
bottom and top is $\exp(2\pi)\approx535$
in a domain $-\pi\leq k_1 z\leq\pi$.
Here, $H_\rho=\cs^2/g$ is the density scale height
and $k_1=2\pi/L$ is the smallest wavenumber that fits into
the cubic domain of size $L^3$.
We adopt Cartesian coordinates $(x,y,z)$,
with periodic boundary conditions in
the $x$ and $y$ directions and stress-free, perfectly conducting
boundaries at top and bottom ($z=\pm L_z/2$).
In most of the calculations, we use a scale separation ratio $\kf/k_1$ of 30,
so $\Gr=0.03$ is still the same as in earlier calculations.
We use a fluid Reynolds number $\Rey\equiv\urms/\nu\kf$ of 36,
and a magnetic Prandtl number $\Pm=\nu/\eta$ of 0.5.
The magnetic Reynolds number is therefore $\Rm=\Pm\Rey=18$.
These values are a compromise between having both $\kf$ and $\Rey$
large enough for NEMPI to develop at an affordable numerical resolution.
The value of $B_0$ is specified in units of $\Beqz=\sqrt{\rho_0} \, \urms$,
where $\rho_0=\bra{\rho}$ is the volume-averaged density,
which is constant in time.
The local equipartition field strength is $\Beq(z)=\sqrt{\rho} \, \urms$.
In our units, $k_1=\cs=\mu_0=\rho_0=1$.
However, time is specified as the turbulent-diffusive time
$t\,\etatz k_1^2$, where $\etatz=\urms/3\kf$ is the estimated
turbulent diffusivity.
We also use DNS to compute these values more accurately with the TFM.
The simulations are performed with the {\sc Pencil Code}
(http://pencil-code.googlecode.com),
which uses sixth-order explicit
finite differences in space and a third-order accurate
time-stepping method.
We use a numerical resolution of $256^3$ mesh points, which was found
to be sufficient for the parameter regime specified above.

\begin{figure*}[t!]\begin{center}
\includegraphics[width=\textwidth]{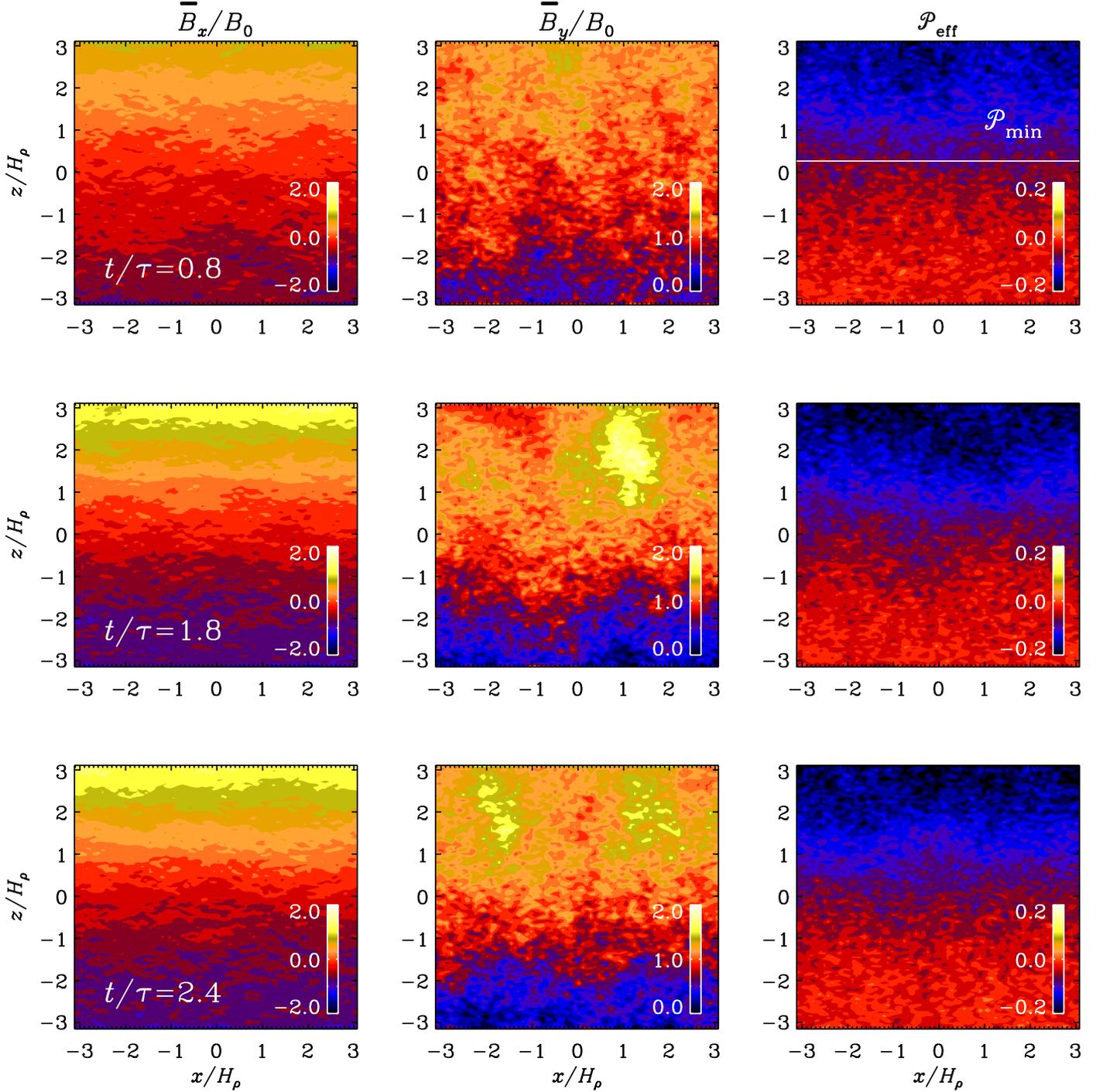}\end{center}\caption[]{
Visualization of $\meanB_x/B_0$ and $\meanB_y/B_0$ together with
effective magnetic pressure for different times.
Here $\Omega=0.15$, $\Co=0.09$, $\Gr=0.033$, and $\kf/k_1=30$.
}\label{DNS_P_Om15_b005_th0b}\end{figure*}

\begin{figure*}[t!]\begin{center}
\includegraphics[width=\textwidth]{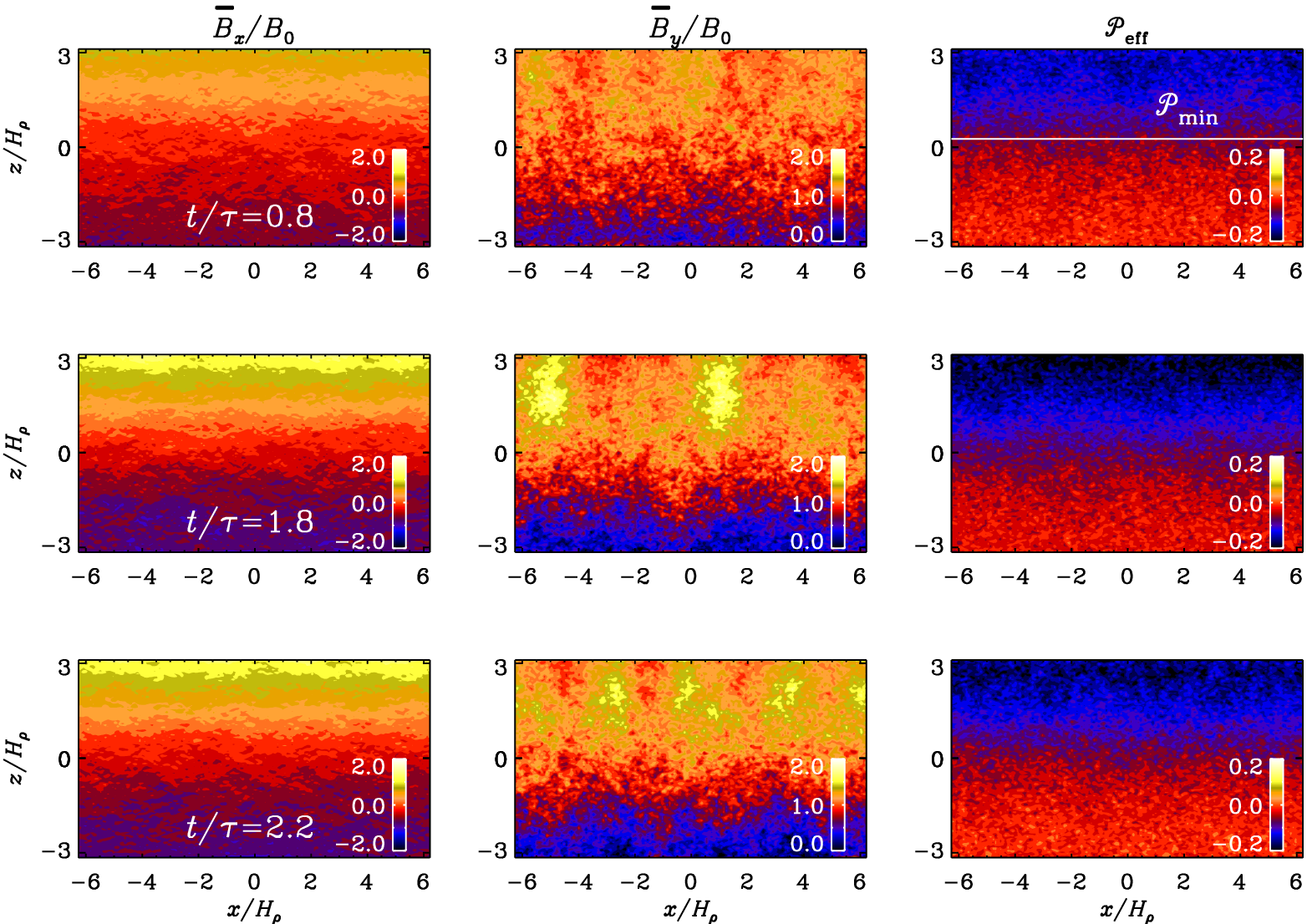}\end{center}\caption[]{
Like \Fig{DNS_P_Om15_b005_th0b}, but for a wider domain.
}\label{DNS_P_Om15_b005_th0b_2pi}\end{figure*}

\begin{figure*}[t!]\begin{center}
\includegraphics[width=\textwidth]{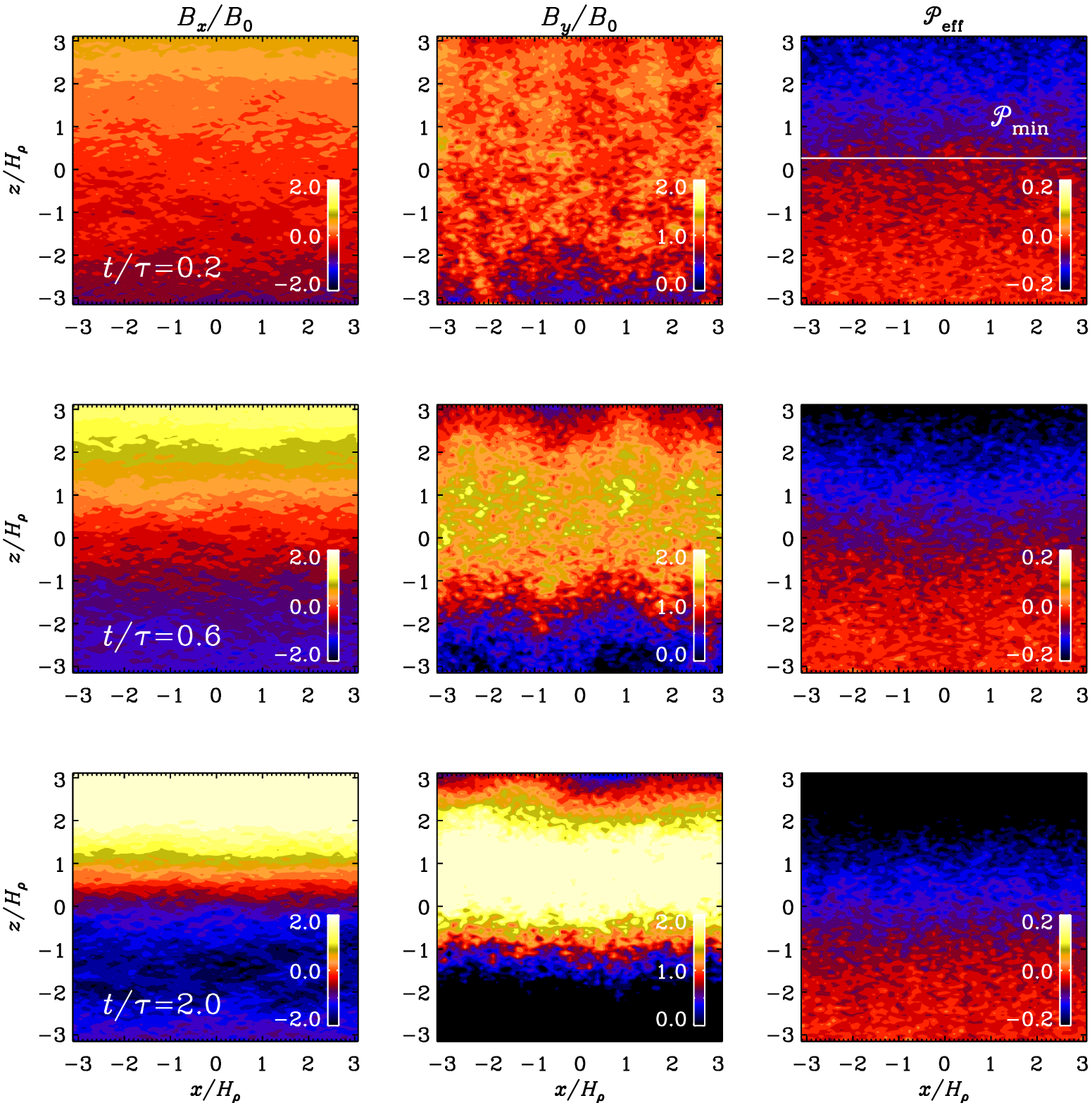}\end{center}\caption[]{
Like \Fig{DNS_P_Om15_b005_th0b}, but for $\Omega=0.35$, so $\Co=0.22$.
}\label{DNS_P_Om35_b005_th0b}\end{figure*}

\subsection{At the verge between NEMPI and dynamo}

The work of \cite{Los2} suggested that for $\Gr=0.03$ and $\Co\ge0.03$,
NEMPI becomes strongly suppressed, and that for still larger values,
the growth rate increases again.
This was tentatively associated with dynamo action, but it was not
investigated in further detail.
We now consider such a case with $\Co=0.09$.
This is a value that resulted in a rather low growth rate for NEMPI, while the
estimated growth rate would be still subcritical for dynamo action.
Following the work of \cite{Los2}, we impose here a horizontal
magnetic field in the $y$ direction with a strength of $0.05\Beqz$,
which was previously found to be in the optimal range for NEMPI to
develop \citep{KBKR12}.

To bring out the structures more clearly, it was found to be advantageous
to present mean magnetic fields by averaging over the $y$ direction
and over a certain time interval $\Delta t$.
We denote such averages by an overbar, e.g., $\meanB_y$.
Once a dynamo develops, we expect a Beltrami-type magnetic field
with $\meanB_x$ phase shifted relative to $\meanB_y$ by $\pi/2$
\citep{B01}.
These are force-free fields with $\nab\times\meanBB=k\meanBB$
such as $\meanBB\propto(\sin kz,\cos kz,0)$, for example.

Figure \ref{DNS_P_Om15_b005_th0b} shows
visualizations of $\meanB_x$ and $\meanB_y$
together with the effective magnetic pressure, $\Peff$ (defined below),
at different times for a value of $\Co$ that is
around the point where we expect onset of dynamo action.
As in earlier work without rotation \citep{KBKMR13},
$\meanB_y$ varies between 0 to $2B_0$.
Furthermore, $\meanB_x$ varies in the range $\pm2B_0$.
In \Fig{DNS_P_Om15_b005_th0b_2pi}, the $x$ extent of the domain is twice
as big: $-2\pi<k_1 x<2\pi$.
In \Fig{DNS_P_Om35_b005_th0b} we show the result for $\Co=0.22$, where
a Beltrami-type field with a $\pi/2$ phase shift between $\meanB_x$ and $\meanB_y$
is well developed.
For smaller values of $\Co$, there are structures (e.g., for $t/\tau=1.8$
at $x/H_\rho\approx1.5$ and for $t/\tau=2.4$
at $x/H_\rho\approx1.5$ and $-2$) that are reminiscent
of those associated with NEMPI.
This can be seen by comparing our \Fig{DNS_P_Om15_b005_th0b} with
Fig.~4 of \cite{KBKMR13} or Fig.~3 of \cite{Los2}.
When the domain is twice as wide, the number of structures simply doubles.
A similar phenomenon was also seen in the simulations of \cite{KeBKMR12}.
For larger values of $\Co$, NEMPI is suppressed and
the $\alpha^2$ dynamo, which generates mean magnetic field of
a Beltrami-type structure, becomes more strongly excited.

The effective magnetic pressure shown in
\Figss{DNS_P_Om15_b005_th0b}{DNS_P_Om35_b005_th0b}
is estimated by computing
the $xx$ component of the total stress from the fluctuating
velocity and magnetic fields as
\EQ
\Delta\overline\Pi_{xx}^{\rm f}
=\meanrho\,(\overline{u_x^2}-\overline{u_{0x}^2})
+\half(\overline{\bb^2}-\overline{\bb_0^2})
-(\overline{b_x^2}-\overline{b_{0x}^2}),
\label{Pi_xx}
\EN
where the subscript $0$ refers to the case with $B_0=0$.
We then calculate \citep{BKKR12}
\EQ
\qp=-2\Delta\overline\Pi_{xx}^{\rm f}/\meanBB^2.
\EN
Here, $\qp(\beta)$ is a function of $\beta=\meanB/\Beq(z)$.
We then calculate $\Peff=\half(1-\qp)\beta^2$, which is the
effective magnetic pressure divided by $\Beq^2$.
We note that $\Peff$ shows a systematic $z$ dependence and is
negative in the upper part.
Variations in the $x$ direction are comparatively weak and
therefore do not show a clear correspondence with the
horizontal variations of $\meanB_y$.

\begin{figure}[t!]\begin{center}
\includegraphics[width=\columnwidth]{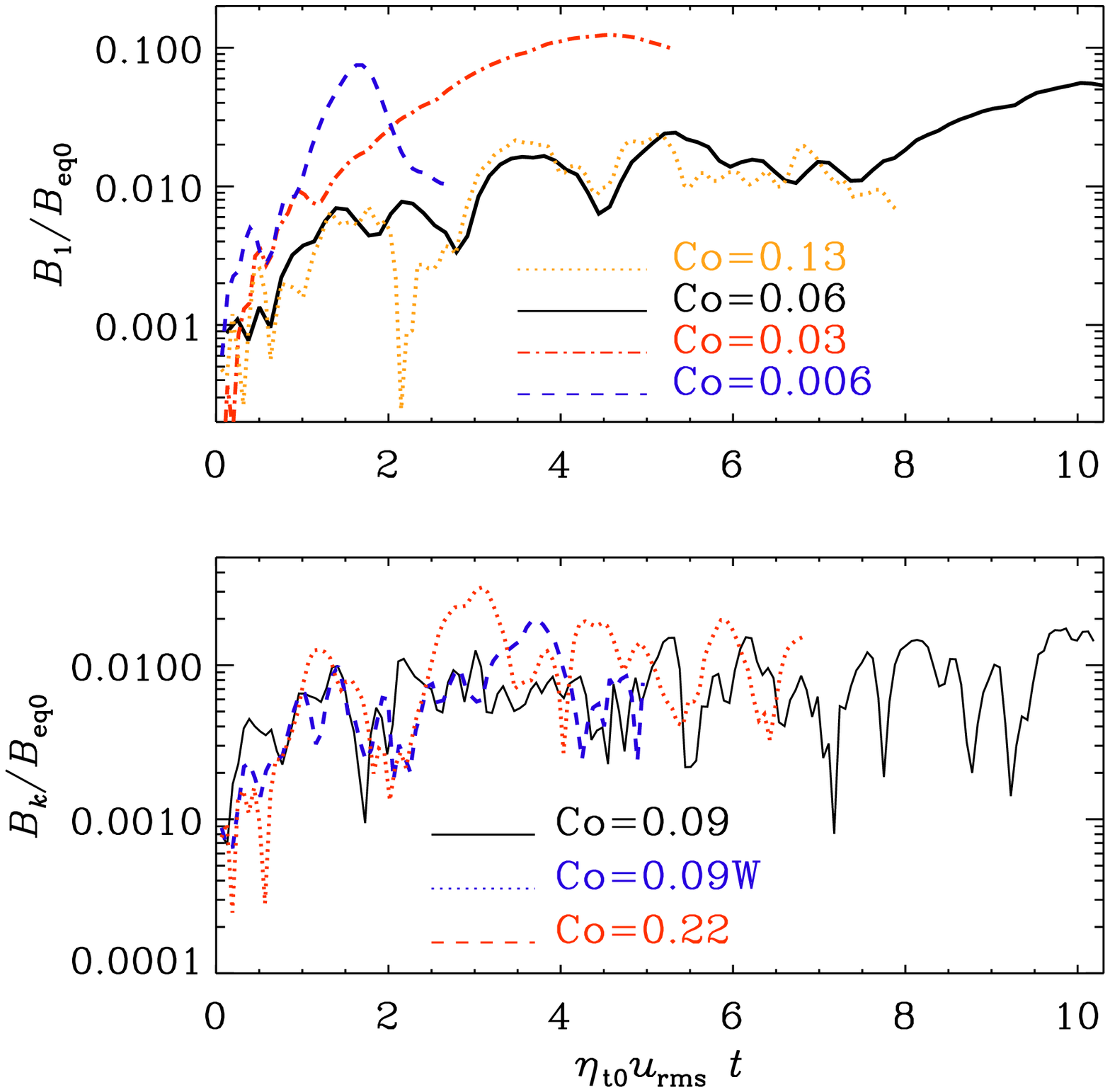}
\end{center}\caption[]{
Comparison of the evolution of $B_k/\Beqz$ for runs with
different values of $\Co$.
In the first panel $k/k_1=1$, while in the second panel
$k/k_1=2$ for the two runs with $\Co=0.09$
(label W refers to the wider box in the $x$ direction),
and $k/k_1=1$ for the run with $\Co=0.22$.
}\label{pB1_vs_t_comp}\end{figure}

As in earlier work \citep{BKKMR11}, we characterize the strength of
resulting structures by an amplitude $B_k$ of a suitably low wavenumber
Fourier mode ($k/k_1=1$ or 2),
which is based on the magnetic field in the upper part of the domain,
$2\leq z/H_\rho\leq\pi$.
In \Fig{pB1_vs_t_comp} we compare the evolution of $B_k/\Beqz$
for runs with different values of $\Co$.
For comparison, we also reproduce the first few runs of \cite{Los2} for $\Co=0.006$--$0.13$,
where we used $k/k_1=1$ in all cases.
It turns out that for the new cases with $\Co=0.09$ and $0.22$, the growth
of $B_k/\Beqz$ is not as strong as for the cases with smaller $\Co$.
Furthermore, as is also evident from
\Figs{DNS_P_Om15_b005_th0b}{DNS_P_Om15_b005_th0b_2pi},
the structures are now characterized by $k/k_1=2$, while for
$\Co=0.22$ they are still characterized by $k/k_1=1$.
The growth for all three cases ($\Co=0.09$, both for normal and wider
domains, as well as $\Co=0.22$) is similar.
However, given that the typical NEMPI structures are not clearly seen for
$\Co=0.22$,  it is possible that the growth
of structures is simply overwhelmed by the much stronger growth due
to the dynamo, which is not reflected in the growth of $B_k/\Beqz$,
whose growth is still mainly indicative of NEMPI.
In this sense, there is some evidence of the occurrence of NEMPI
in both cases.

\subsection{Kinetic helicity}

\begin{figure}[t!]\begin{center}
\includegraphics[width=\columnwidth]{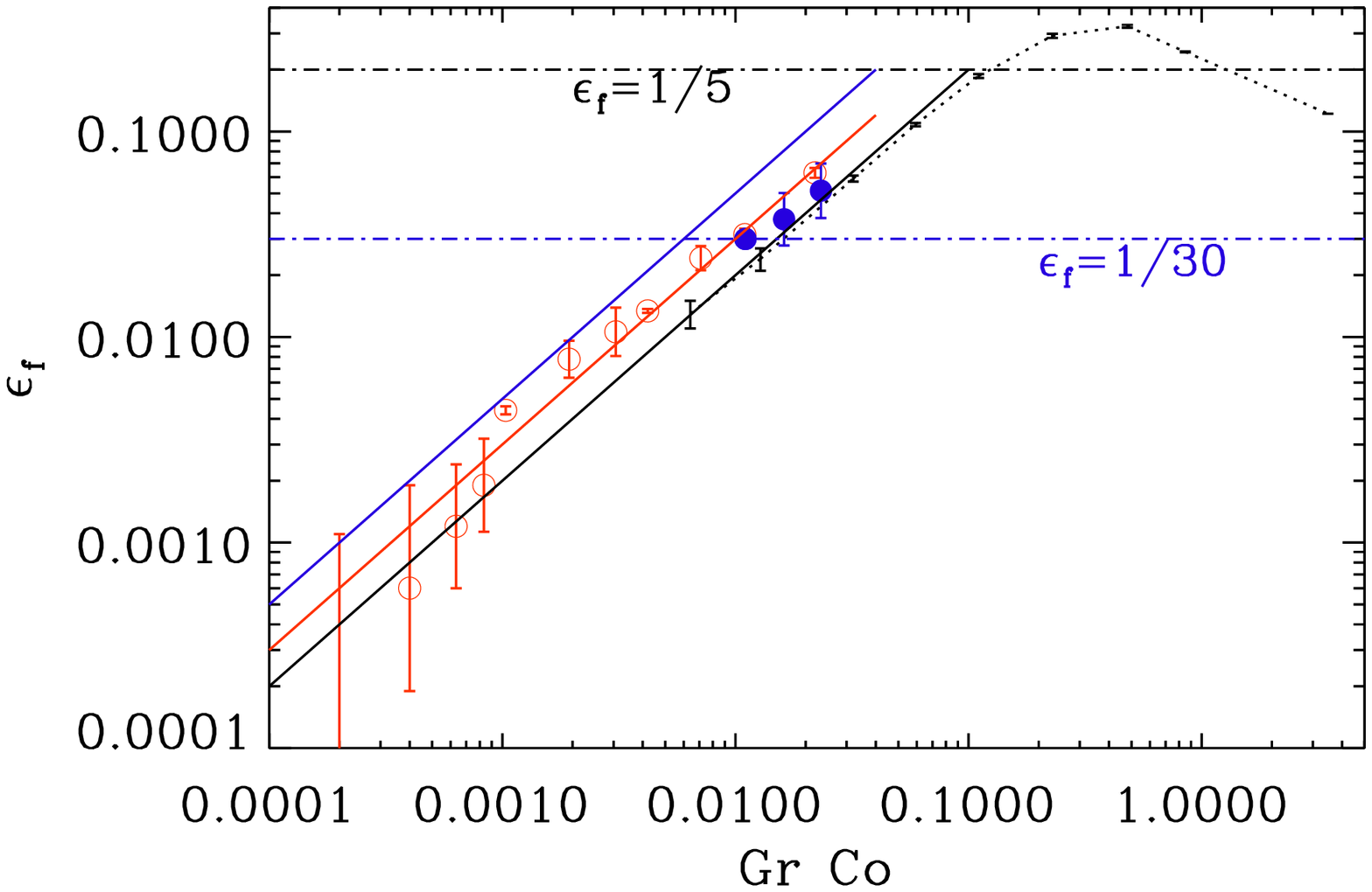}
\end{center}\caption[]{
Dependence of $\epsf$ on $\Gr\,\Co$ obtained in DNS with imposed field
(open symbols, red) and without (closed symbols, blue), for $\kf/k_1=30$.
The black symbols connected by a dotted line correspond to the
values of \cite{BRK12} for $\kf/k_1=5$.
The horizontal lines correspond to the dynamo threshold
for the two values of $\kf/k_1$.
}\label{poum_Odep_g08}
\end{figure}

To estimate the $\alpha$ effect and study its relation to kinetic helicity
we begin by considering a fixed value of $\Gr$ equal to that
used by \cite{Los2} and vary $\Co$.
For small values of $\Co$, their data agreed with the MFS of \cite{Los1}.
For faster rotation, one eventually crosses the dynamo threshold.
This is also the point at which the growth rate begins to increase again,
although it now belongs to a different instability than
for small values of $\Co$.
The underlying mechanism is the $\alpha^2$-dynamo, which is characterized
by the dynamo number
\EQ
C_\alpha=\alpha/\etaT k_1,
\label{Calp}
\EN
where $\alpha$ is the typical value of the $\alpha$ effect (here assumed
spatially constant), $\etaT=\etat+\eta$ is the sum of turbulent
and microphysical magnetic diffusivities, and $k_1$ is the
lowest wavenumber of the magnetic field that can be fitted
into the domain.
For isotropic turbulence, $\alpha$ and $\etat$ are respectively
proportional to the negative kinetic helicity and the
mean squared velocity \citep{Mof78,KR80,RKR03,KR03}
\EQ
\alpha\approx\alpha_0\equiv-\onethird\tau\overline{\oo\cdot\uu},\quad
\etat\approx\etatz\equiv\onethird\tau\overline{\uu^2},
\label{alp0_etatz}
\EN
where $\tau=(\urms\kf)^{-1}$, so that \citep{BB02,CB13}
\EQ
C_\alpha=-\epsk\epsf\kf/k_1.
\label{Calp0}
\EN
Here, $\epsk$ is a free parameter characterizing
possible dependencies on the forcing wavenumber,
and $\epsf$ is a measure for the relative kinetic helicity.
Simulations of \cite{BRK12} and \cite{Los2} showed that
\EQ
\epsf\equiv\overline{\oo\cdot\uu}/\kf\urms^2
\approx\epsfz\,\Gr\,\Co\quad(\Gr\,\Co\la0.1),
\label{epsf}
\EN
where $\epsfz$ is yet another non-dimensional parameter on the order of unity
that may depend weakly on the scale separation ratio, $\kf/k_1$,
and is slightly different with and without imposed field.
In the absence of an imposed field, \cite{BRK12} found $\epsfz\approx2$
using $\kf/k_1=5$.
However, both an imposed field and a larger value of $\kf/k_1$ lead to a
slightly increased value of $\epsfz$.
Our results are summarized in \Fig{poum_Odep_g08} for cases with
and without imposed magnetic fields.
Error bars are estimated as the largest departure of any one third
of the full time series.
The relevant points of \cite{Los2} give $\epsfz\approx2.8$.
For $\Gr\,\Co\ga0.5$, the results of \cite{BRK12} show a maximum
with a subsequent decline of $\epsf$ with increasing values of $\Co$.
However, although it is possible that the position of this maximum may
be different for other values of $\Gr$, it is unlikely to be relevant
to our present study where we focus on smaller values of $C_\alpha$
near dynamo onset.
Thus, in conclusion, \Eq{epsf} seems to be a useful approximation that
has now been verified over a range of different values of $\kf/k_1$.

\subsection{Test-field results}

Our estimate for $C_\alpha$ is based on the reference values $\alpha_0$ and
$\etatz$ that are defined in \Eq{alp0_etatz} and represent approximations
obtained from earlier simulations of helically forced turbulence \citep{SBS08}.
In the present study, helicity is self-consistently generated from
the interaction between rotation and stratification.
As an independent way of computing $\alpha$ and $\etat$, we now use
the test-field method (TFM).
It consists of solving auxiliary equations describing the evolution of
magnetic fluctuations, $\bb^{pq}$, resulting from a set of several prescribed
mean or test fields, $\meanBB^{pq}$.
We solve for the corresponding vector potential $\aaaa^{pq}$ with
$\bb^{pq}=\nab\times\aaaa^{pq}$,
\EQ
{\partial\aaaa^{pq}\over\partial t}
=\uu\times\meanBB^{pq}
+\meanUU\times\bb^{pq}
+(\uu\times\bb^{pq})'
+\eta\nabla^2\aaaa^{pq},
\EN
where $(\uu\times\bb^{pq})'=\uu\times\bb^{pq}-\overline{\uu\times\bb^{pq}}$
is the fluctuating part of the electromotive force and
\EQ
\meanBB^{i{\rm c}}=\xxx_i\cos kz,\quad
\meanBB^{i{\rm s}}=\xxx_i\sin kz,\quad
i=1,2,
\EN
are the four test fields, which can show a cosine or sine variation
with $z$, while $\xxx_1=(1,0,0)$ and $\xxx_2=(0,1,0)$ are unit vectors in
the two horizontal coordinate directions.
The resulting $\bb^{pq}$ are used to compute the electromotive force,
$\meanEMF^{pq}=\overline{\uu\times\bb^{pq}}$, which is then expressed
in terms of $\meanBB^{pq}$ and $\meanJJ^{pq}=\nab\times\meanBB^{pq}$ as
\EQ
\meanemf^{pq}_i=\alpha_{ij}\meanB^{pq}_j-\eta_{ij}\meanJ^{pq}_j.
\EN
By doing this for all four test field vectors, the $x$ and $y$ components
of each of them gives eight equations for the eight unknowns, $\alpha_{11}$,
$\alpha_{12}$, ..., $\eta_{22}$ \citep[for details see][]{B05QPO}.

With the TFM, we obtain the kernels $\alpha_{ij}$ and $\eta_{ij}$,
from which we compute
\EQ
\alpha=\half(\alpha_{11}+\alpha_{22}),\quad
\etat=\half(\eta_{11}+\eta_{22}),
\EN
\EQ
\gamma=\half(\alpha_{21}-\alpha_{12}),\quad
\delta=\half(\eta_{21}-\eta_{12}) .
\EN
We normalize $\alpha$ and $\etat$ by their respective values
obtained for large magnetic Reynolds numbers defined in \Eq{alp0_etatz},
and denote them by a tilde, i.e., $\tilde\alpha=\alpha/\alpha_0$ and
$\tilde\etat=\etat/\etatz$.
We use the latter normalization also for $\delta$, i.e.,
$\tilde\delta=\delta/\etatz$, but expect its value to vanish in the
limit of zero angular velocity.
No standard turbulent pumping velocity is expected \citep{KR80,Mof78}, because
the rms turbulent velocity is independent of height.
However, this is not quite true.
To show this, we normalize $\gamma$ by $\urms$ and present
$\tilde\gamma=\gamma/\urms$.
In our normalization, the molecular value is given by

$\eta/\eta_0=3/\Rm$.
\begin{figure}[t!]\begin{center}
\includegraphics[width=\columnwidth]{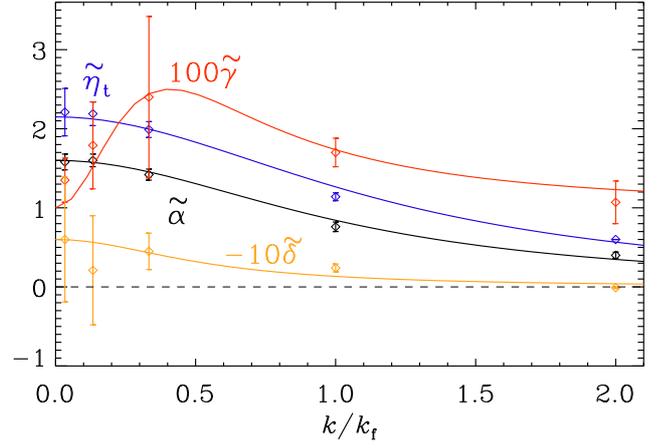}
\end{center}\caption[]{
TFM coefficients versus scale separation ratio, $k/\kf$, for
$\Co=0.59$, $\Rm=18$, $B_{0y}/\Beqz=0.05$,
$\tilde{g}=1$, and $\eta k_1/\cs=2\times10^{-4}$.}
\label{testk}\end{figure}

\begin{figure}[t!]\begin{center}
\includegraphics[width=\columnwidth]{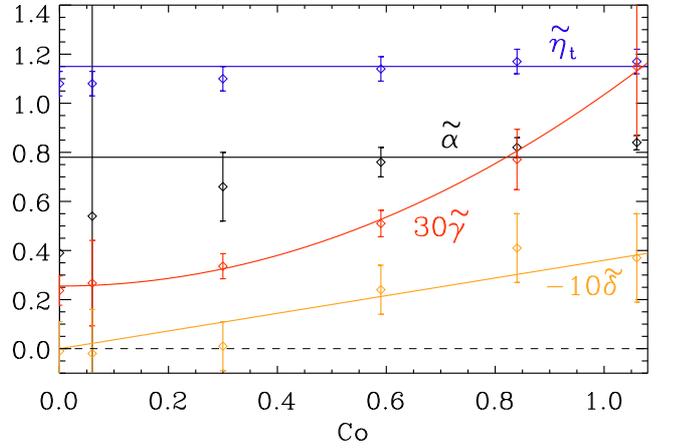}
\end{center}\caption[]{
TFM coefficients versus Coriolis number, $\Co$,
for $k/k_1=1$, $\Rm=18$, $B_{0y}/\Beqz=0.05$,
$\tilde{g}=1$, and $\eta k_1/\cs=2\times10^{-4}$.
}\label{testCo}\end{figure}

We consider test fields that are constant in time and vary sinusoidally
in the $z$ direction.
We choose certain values of $k$ between $k_1$ and $60\kf$ (=$2k_1)$ and
also vary the value of $\Co$ between 0 and about 1.06
while keeping $\Gr=0.033$ fixed.
In all cases where the scale separation ratio is held fixed,
we used $\kf/k_1\approx30$,
which is larger than what has been used in earlier studies \citep{BRS08},
where $\kf/k_1$ was typically 5.

In Figure \ref{testk} we show the dependence of the coefficients
on the normalized wavenumber of the test field, $k/\kf$.
The three coefficients $\tilde\alpha$, $\tilde\etat$, and $\tilde\delta$
show the same behavior of the form of
\EQ
\tilde\sigma=\tilde\sigma_{0}/\left(1+\ell_\sigma^2k^{2}\right)
\EN
for $\tilde\sigma=\tilde\alpha$, $\tilde\etat$, or $\tilde\delta$,
while for $\tilde\gamma$ we use
\EQ
\tilde\gamma=\tilde\gamma_{0}+\tilde\gamma_{2}\ell_{\gamma}^2k^{2}/
\left(1+\ell_\gamma^2k^{2}\right),
\EN
where $\tilde\gamma_{0}=0.01$, $\tilde\gamma_{2}=0.06$, and $\ell_{\gamma}=2.5$.
These results have been obtained for $\Co=0.59$ and $B_{0y}/\Beqz=0.05$.
Again, error bars are estimated as the largest departure of any one third
of the full time series.

Most of the coefficients are only weakly dependent on the
value of $\Co$, except $\gamma$ and $\delta$.
The former varies approximately as
\EQ
\tilde\gamma=\tilde\gamma_{0}+\tilde\gamma_{\Omega}^{2}\Co^2,
\EN
where $\tilde\gamma_{0}=0.85$ and $\tilde\gamma_{\Omega}^{2}=2.6$.
Here and in the following, we keep $k/\kf=1/30$.
For the same value of $k/\kf$, the functional form for $\delta$ shows a linear
increase with $\Co$, i.e., $\tilde\delta=\tilde\delta_{0}\Co$
where $\tilde\delta_{0}=0.036$.
Figure~\ref{testCo} shows that $\tilde\alpha$ is nearly independent of
the Coriolis number.
This result is in agreement with that obtained \cite{KR03}, where
a theory of $\alpha$ versus Coriolis number was developed for large
fluid and magnetic Reynolds numbers.
It turns out that the new values of $\alpha$ and $\etat$
that have been obtained now with the TFM are somewhat
different from previous TFM studies that originally estimated
($\tilde\alpha\approx0.8$ and $\tilde\etat\approx1.15$).
The TFM results now suggest $\epsk=0.6$ in \Eq{Calp0}.
The reason for the departure from unity cannot just be the fact that helicity
is now self-consistently generated, because this was also the case in the
earlier work of \cite{BRK12}.
The only plausible reason is the large value of
$\kf/k_1$ that is now much larger than before
($30$ compared to $5$ in most previous studies), which explains
the reason for our choice of the subscript in $\epsk$.

The origin of weak pumping found in \Figs{testk}{testCo} is unclear.
For a weak mean magnetic field, pumping of the magnetic field can cause
not only inhomogeneous distributions of the velocity fluctuations
\citep{KR80,Mof78} or magnetic fluctuations \citep{RKR03},
but also non-uniform distribution of the fluid density
in the presence either of small-scale dynamo or turbulent convection
\citep{RK06}.
In our simulations there is no small-scale dynamo effect,
because $\Rm$ is too low.
There is also no turbulent convection possible in our setup.
The pumping effect is also not connected with
nonlinear effects; see Fig.~2 in \cite{RK04}.

\section{MFS study}

We now want to see whether the suppression of NEMPI and the subsequent
increase in the resulting growth rate can be reproduced in MFS.
In addition to a parameterization for the negative effective magnetic
pressure in the momentum equation, we add one for the electromotive force.
The important terms here are the $\alpha$ effect and the turbulent
magnetic diffusivity, whose combined effect is captured by the quantity
$C_\alpha$, which is defined in \Eq{Calp} and related to DNS parameters
in \Eq{Calp0}.
In contrast to DNS, the advantage of MFS is that they can more easily
be extended to astrophysically interesting conditions of large Reynolds
numbers and more complex geometries.

\subsection{The model}

Our MFS model is in many ways the same as that of \cite{Jab1},
where parameterizations for negative effective magnetic pressure
and electromotive force where, for the first time, considered
in combination with each other.
Their calculations were performed in spherical shells without
Coriolis force, while here we apply instead Cartesian geometry
and do include the Coriolis force.
The evolution equations for mean velocity $\meanUU$,
mean vector potential $\meanAA$, and mean density $\meanrho$, are thus
\EQA
{\DD\meanUU\over\DD t}\!\!&=&\!\!{1\over\meanrho}\left(
\meanJJ\times\meanBB+\nab{q_{\rm p}\meanBB^2\over2}\right)
-2\OO\times\meanUU-\nuT\meanQQ+\meanFF,\\
\label{dAmean}
{\partial\meanAA\over\partial t}\!&=&\!\meanUU\times\meanBB
+\alpha\meanBB-\etaT\meanJJ,\\
{\DD\meanrho\over\DD t}\!&=&\!-\meanrho\nab\cdot\meanUU,
\nonumber
\ENA
where $\DD/\DD t=\partial/\partial t+\meanUU\cdot\nab$
is the advective derivative,
\EQ
\meanFF=\grav-\cs^2\nab\ln\meanrho
\EN
is the mean-field hydrostatic force balance,
$\etaT=\etat+\eta$ and $\nuT=\nut+\nu$ are the sums of turbulent and
microphysical values of magnetic diffusivity and kinematic viscosities,
respectively, $\alpha$ is the aforementioned coefficient
in the $\alpha$ effect,
$\meanJJ=\nab\times\meanBB$  is the mean current density,
\EQ
-\meanQQ=\nabla^2\meanUU+\onethird\nab\nab\cdot\meanUU
+2\meanSSSS\nab\ln\meanrho
\EN
is a term appearing in the viscous force, where
$\meanSSSS$ is the traceless rate of strain tensor of the mean flow
with components $\meanSSS_{ij}=\half(\meanU_{i,j}+\meanU_{j,i})
-\onethird\delta_{ij}\nab\cdot\meanUU$,
and finally $\nab(q_{\rm p}\meanBB^2/2)$
determines the turbulent contribution to the mean Lorentz force.
Here, $q_{\rm p}$ depends on the local field strength
and is approximated by \citep{KBKR12}
\EQ
\qp(\beta)={\qpz\over1+\beta^2/\betap^2}
={\betastar^2\over\betap^2+\beta^2},
\label{qp-apr}
\EN
where $\qpz$, $\betap$, and $\betastar=\betap\qpz^{1/2}$ are constants,
$\beta=|\meanBB|/\Beq$ is the normalized mean
magnetic field, and $\Beq=\sqrt{\rho}\, \urms$
is the equipartition field strength.
For $\Rm\la60$, \cite{BKKR12} found $\betastar\approx0.33$
and $\betap\approx1.05/\Rm$.
We use as our reference model the parameters for $\Rm=18$,
also used by \cite{Los2}, which yields
\EQ
\betap=0.058,\quad\betastar=0.33
\quad\mbox{(reference model)}.
\label{ModelParam}
\EN
In some cases we also compare with $\betastar=0.44$, which was found
to match more closely the measured dependence of the effective
magnetic pressure on $\beta$ by \cite{Los2}.
For vertical magnetic fields, MFS for a range of model parameters
have been given by \cite{Jab2}.
In the MFS, we use \citep{SBS08}
\EQ
\etat\approx\etatz\equiv\urms/3\kf
\label{etat}
\EN
to replace $\kf=\urms/3\etat$, so
\EQ
\Gr=3\etat/\urms H_\rho
\EN
and \citep{Los2}
\EQ
\Co=2\Omega/\urms\kf=6\Omega\etat/\urms^2.
\EN
We now consider separately cases where we vary either $\Co$ or $\Gr$.
In addition, we also vary the scale separation ratio $\kf/k_1$, which
is essentially a measure of the inverse turbulent diffusivity, i.e.,
\EQ
\kf/k_1=\urms/3\etat k_1
\EN
(see \Eq{etat}).

\subsection{Fixed value of $\Gr$}

The work of \cite{Los1} has shown that the growth rate of NEMPI,
$\lambda$, decreases with increasing values of the rotation rate.
They found it advantageous to express $\lambda$ in terms of the quantity
\EQ
\lambda_{*0}=\betastar\urms/H_\rho.
\label{lamstar}
\EN
As discussed above,
the normalized growth rate $\lambda/\lambda_{*0}$ shows first a decline with
increasing values of $\Co$, but then an increase for $\Co>0.13$,
which was argued to be a result of the dynamo effect \citep{Los2}.
This curve has a minimum at $\Co\approx0.13$.
As rotation is increased further, the combined action of stratification
and rotation leads to increased kinetic helicity and thus eventually
to the onset of mean-field $\alpha^2$ dynamo action.

\begin{figure}[t!]\begin{center}
\includegraphics[width=\columnwidth]{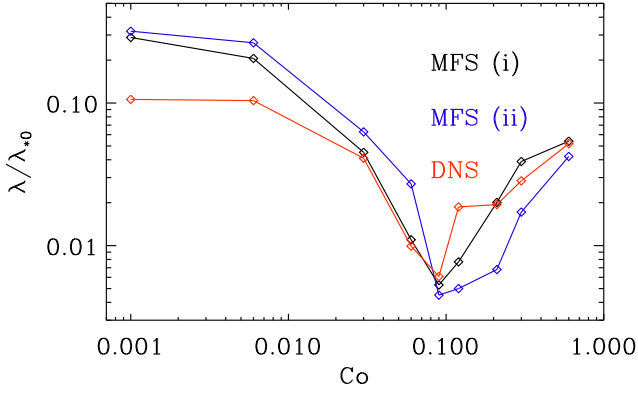}
\end{center}\caption[]{Non-dimensional
growth rate of NEMPI versus $\Co$ for MFS(i) with
$\betastar=0.33$ and MFS(ii) with $\betastar=0.44$, as well as
DNS for $\Gr=0.033$ and $\beta_0=0.05$.
}\label{lambda}\end{figure}

Owing to the effects of turbulent diffusion, the actual value of the
growth rate of NEMPI is always expected to be less than $\lambda_{*0}$.
\cite{KBKMR13} proposed an empirical formula replacing $\lambda$ by
$\lambda+\etat k^2$, where $k$ is the wavenumber.
This would lead to
\EQ
\lambda/\lambda_{*0}\propto1-\Gr_*/\Gr,
\label{lamlam0old}
\EN
with a coefficient $\Gr_*=\tilde\etat/3\betastar\Ma$.
However, as we will see below, this expression is not found to
be consistent with our numerical data.

The onset of the dynamo instability is governed by the dynamo number
\EQ
C_\alpha=\epsfz\,\Gr\,\Co\,\kf/k_1.
\label{Calp1}
\EN
For a cubic domain, large-scale dynamo action occurs for $C_\alpha>1$,
which was confirmed by \cite{Los2}, who found the typical Beltrami
fields for two supercritical cases.
They used the parameters $\Gr=0.033$ and values of $\Co$ up to 0.6.
Here we present MFS in two- and three-dimensional domains
for the same values of $\Gr$ and a similar range of $\Co$ values.
In \Fig{lambda}, we compare the DNS of \cite{Los2} with our reference model
defined through \Eq{ModelParam} and referred to as MFS(i) as well as with the
case $\betastar=0.44$, referred to as MFS(ii).

\begin{figure}[t!]\begin{center}
\includegraphics[width=\columnwidth]{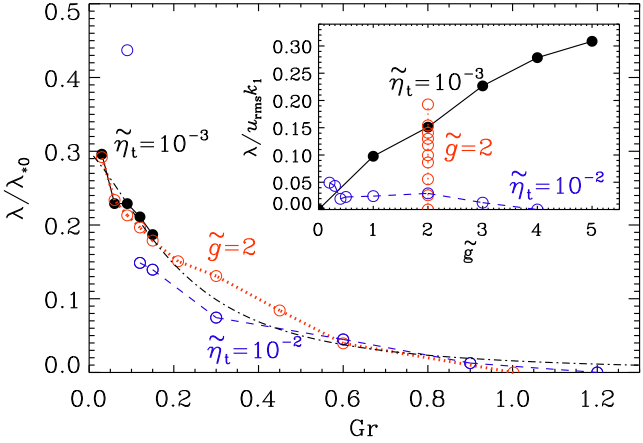}
\end{center}\caption[]{
Normalized growth rate of NEMPI versus stratification parameter $\Gr$ that
varies with changing gravity, $g$, for $\Co=0$ with constant
$\tilde\etat$ ($\tilde\etat=10^{-3}$ black filled symbols
and $\tilde\etat=10^{-2}$ blue open symbols),
or with changing $\etat=\nut$ for constant
$\tilde{g}=2$ (red open symbols).
The dash-dotted line shows the approximate fit given by \Eq{lamGrfit}.
The inset shows the growth rate normalized by the turnover time
as a function of $\tilde{g}$.
}\label{presults_Co0}\end{figure}

\subsection{Larger stratification, smaller scale separation}

The expected theoretical maximum growth rate of NEMPI is given by
\Eq{lamstar}.
At zero rotation, we thus expect $\lambda/\lambda_{*0}\approx1$.
To check this, we performed two-dimensional MFS in a squared domain
of size $(2\pi)^2$.
The result is shown in \Fig{presults_Co0} for the model parameters
given in \Eq{ModelParam}.
When $\Gr$ is small, we find that $\lambda/\lambda_{*0}\approx0.3$,
which is below the expected value.
As we increase $\Gr$, $\lambda/\lambda_{*0}$ decreases until NEMPI
can no longer be detected for $\Gr\ga1.2$.

It is conceivable that this decrease may have been caused by
the following two factors.
First, the growth rate is expected to increase with $\Gr$, but
for fixed scale separation, the resulting density contrast becomes huge.
Finite resolution might therefore have caused inaccuracies.
Second, although the growth rate should not depend on $B_0$
\citep{KBKR12},
we need to make sure that the mode is fully contained within the domain.
In other words, we are interested in the largest growth rate
as we vary the value of $B_0$.
Again, to limit computational expense, we tried only a small
number of runs, keeping the size of the domain the same.
This may have caused additional uncertainties.
However, it turns out that our results are independent of whether
$\Gr$ is changed by changing $g$ or $\etat$ ($=\nut$).
This suggests that our results for large values of $g$ shown in
\Fig{presults_Co0} may in fact be accurate.
To illustrate this more clearly, we rewrite
\EQ
\Gr={3\etat\over\urms H_\rho}
={3\tilde\etat\over k_1}\,{\cs\over\urms}\,{g\over\cs^2}
=3\tilde\etat\tilde{g}/\Ma,
\label{Gr_etat_and_g}
\EN
where we have defined
\EQ
\tilde\etat=\etat k_1/\cs,\quad
\tilde{g}=g/\cs^2 k_1\equiv(k_1 H_\rho)^{-1}.
\EN
\FFig{presults_Co0} shows that $\lambda/\lambda_{*0}$ is indeed independent
of the individual values of $\tilde\etat$ and $\tilde{g}$ as long as
$\Gr$ is the same.
For small values of $\tilde{g}$ and
large diffusivity ($\tilde\etat=10^{-2}$), the velocity
evolves in an oscillatory fashion with a rapid growth
and a gradual subsequent decline.
In \Fig{presults_Co0}, the isolated data point at
$\lambda/\lambda_{*0}\approx0.44$ reflects the speed of growth
during the periodic rise phase, but it is unclear whether or not
it is related to NEMPI.

In the inset, we plot $\lambda/\urms\kf$ versus $\tilde{g}$ itself.
This shows that the growth rate (in units of the inverse turnover time)
increases with increasing $\tilde{g}$ when $\tilde\etat$ is small.
However, as follows from theoretical analysis,
the growth rate decreases with increasing $\tilde\etat$.
When $\tilde\etat$ is larger (corresponding to smaller scale separation),
the growth rate of NEMPI is reduced for the same value of $\tilde{g}$ and
it decreases with $\tilde{g}$ when $\tilde{g}\ga2$.

The decrease of $\lambda/\lambda_{*0}$ with increasing values of $\Gr$
can be approximated by the formula
\EQ
\lambda/\lambda_{*0}\approx0.3\left/\left[1+2\Gr+(4\Gr)^2\right]\right.,
\label{lamGrfit}
\EN
which is shown in \Fig{presults_Co0} as a dash-dotted line.
This expression is qualitatively different from the earlier,
more heuristic expression proposed by \cite{KBKMR13} where
the dimensional growth rate was simply modified by an ad hoc
diffusion term of the form $\etat k^2$.
In that case, contrary to our MFS,
the normalized growth rate would actually
increase with increasing values of $\Gr$ (see \Eq{lamlam0old}).

\subsection{$\Co$ dependence at larger stratification}

\begin{figure}[t!]\begin{center}
\includegraphics[width=\columnwidth]{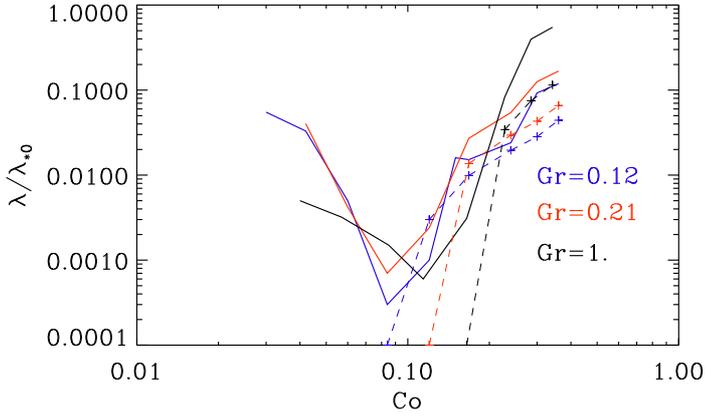}
\end{center}\caption[]{
Normalized growth rate of the combined NEMPI and dynamo instability
(solid lines) together with cases with pure dynamo instability
(no imposed field, dashed lines)
versus $\Co$ for three different values of $\Gr$; $\Gr=0.12$ (blue),
$\Gr=0.21$ (red), and $\Gr=1.0$ (black).
In these simulations $\tilde{g}=4$ and $\tilde\etat=10^{-3}$
(blue line), $\tilde{g}=3.5$, $\tilde\etat=2\times10^{-3}$ (red line),
and $\tilde{g}=3.5$, $\tilde\etat=9.5\times10^{-3}$ (black line).
}\label{presults_Gr_comp}\end{figure}

We consider the normalized growth rate of the
combined NEMPI and dynamo instabilities as a function of $\Co$ for
different values of $\Gr$.
As is clear from \Fig{presults_Co0}, using a fixed value of $g$
and varying $\etat$ gives us
the possibility to increase $\Gr$ to larger values of up to 1.
In the following we use this procedure to compare the behavior of the
growth rate versus $\Co$ for three values of $\Gr$, $0.12$,
$0.21$, and $1$ (see \Fig{presults_Gr_comp}).
It can be seen that the behavior of the curves is independent of the values
of $\Gr$, but the points where the minima of the curves occur moves toward
bigger values of $\Co$ as $\Gr$ increases.
This also happens in the case when there is only dynamo action without
imposed magnetic field (dashed lines in \Fig{presults_Gr_comp}).
One also sees that the increase of the growth rate with increasing $\Co$
is much stronger in the case of larger $\Gr$
(compare the lines for $\Gr=0.12$ with those for 0.21 and 1).
Finally, comparing runs with and without imposed magnetic field, but
the same value of Gr, the growth rate of NEMPI is in most cases below
that of the coupled system with NEMPI and dynamo instability.

In \Fig{presults_Gr_comp} we see that the dependence
of $\lambda/\lambda_{*0}$ on $\Gr$ is opposite for
small and large values of $\Co$.
When $\Co\la0.05$, an increase in $\Gr$ leads to a
decrease in $\lambda/\lambda_{*0}$ (compare the $\Gr=1$ line with that for
0.21 along a cut through $\Co=0.05$ in \Fig{presults_Gr_comp}), while for
$\Co\ga0.2$, an increase in $\Gr$ leads to an
increase in $\lambda/\lambda_{*0}$ (compare all three lines
in \Fig{presults_Gr_comp} along a cut through $\Co=0.3$).
The second case is caused by the increase of the dynamo
number $C_\alpha$, which is directly proportional
to $\Gr$ (see \Eq{Calp1}).
On the other hand, for small values of $\Co$, only NEMPI operates, but
if $\Gr$ in \Eq{Gr_etat_and_g} is increased by increasing $\tilde\etat$
rather than $\tilde{g}$, the dynamo is suppressed by enhanced
turbulent diffusion (see also \Fig{presults_Co0}).
This is related to the fact that the properties of the system
depend not just on $\Gr$ and $\Co$, but also on $\kf/k_1$ or
$C_\alpha$, which is proportional to all three parameters
(see \Eq{Calp1}).

\section{Discussion and conclusions}

The present work has brought us one step closer to being able
to determine whether
the observable solar activity such as sunspots and active regions could
be the result of surface effects associated with strong stratification.
A particularly important aspect has been the interaction with a dynamo
process that must ultimately be responsible for generating the overall
magnetic field.
Recent global convective dynamo simulations of \cite{NBBMT11,NBBMT13,NBBMT14}
have demonstrated that flux tubes with $\approx40\kG$ field strength can
be produced in the solar convection zone.
This is almost as strong as the $\approx100\kG$ magnetic flux tubes
anticipated from earlier investigations of rising flux tubes requiring
them to not break up and to preserve their east--west orientation
\citep{DSC93}.
Would we then still need surface effects such as NEMPI to produce
sunspots?
The answer might well be yes, because the flux ropes that have been
isolated in the visualizations of \cite{NBBMT11,NBBMT13,NBBMT14} appear to
have cross sections that are much larger than sunspots at the solar surface.
Further concentration into thinner tubes would be required if they
were to explain sunspots by just letting them pierce the surface.

Realistic hydromagnetic simulations of the solar surface are now beginning
to demonstrate that $\approx10\kG$ fields at a depth of $\approx10\Mm$
can produce sunspot-like appearances at the surface \citep{RC14}.
However, we have to ask about the physical process contributing to this
phenomenon.
A purely descriptive analysis of simulation data cannot replace the need
for a more prognostic approach that tries to reproduce the essential
physics using simpler models.
Although \cite{RC14} propose a mechanism involving mean-field terms in
the induction equation, they do not show that their model equations
can actually describe the process of magnetic flux concentration.
In fact, their description is somewhat reminiscent of flux expulsion,
which was invoked earlier by \cite{TWBP98} to explain the segregation
of magneto-convection into magnetized and unmagnetized regions.
In this context, NEMPI provides such an approach that can be used
prognostically rather than diagnostically.
However, this approach has problems of its own, some of which are
addressed in the present work.
Does NEMPI stop working when $\Co\ga0.03$?
How does it interact with the underlying dynamo?
Such a dynamo is believed to control the overall sunspot number and
the concentration of sunspots to low latitudes.

Our new DNS suggest that, although rotation tends to suppress NEMPI,
magnetic flux concentrations can still form at Coriolis numbers
of $\Co\approx0.1$.
This is slightly larger than what was previously found from MFS
both with horizontal and vertical magnetic fields and the same
value of $\Gr$.
For the solar rotation rate of $\Omega\approx3\times10^{-6}\s^{-1}$,
a value of $\Co\equiv2\Omega\tau=0.1$ corresponds to $\tau=5\h$, which
is longer than the earlier MFS values of $2\h$ for a horizontal
field \citep{Los2} and $30\min$ for a vertical field \citep{Jab2}.

Using the TFM, we have confirmed earlier findings regarding $\alpha$ and
$\etat$, although for our new simulations both coefficients are somewhat
larger, which is presumably due to the larger scale separation.
The ratio between $\alpha$ and $\etat$
determines the dynamo number and is now about 40\% below
previous estimates.
There is no evidence of other important mean-field effects that
could change our conclusion about a cross-over from suppressed NEMPI
to increased dynamo activity.
We now confirm quantitatively that the enhanced growth past
the initial suppression of NEMPI is indeed caused by mean-field
dynamo action in the presence of a weak magnetic field.
The position of the minimum in the growth rate coincides with
the onset of mean-field dynamo action that takes the $\alpha$ effect
into account.

For weak or no rotation, we find that the normalized NEMPI growth rate
is described by a single parameter $\Gr$, which is proportional to
the product of gravity and turbulent diffusivity, where the latter
is a measure of the inverse scale separation ratio.
This normalization takes into account that the growth rate increases
with increasing gravity.
The growth rate compensated in this way shows a decrease with
increasing gravity and turbulent diffusivity that is different
from an earlier, more heuristic, expression proposed by
\cite{KBKMR13}.
The reason for this departure is not quite clear.
One possibility is some kind of gravitational quenching, because
the suppression is well described by a quenching factor that becomes
important when $\Gr$ exceeds a value of around 0.5.
This quenching is probably not important for stellar convection
where the estimated value of $\Gr$ is 0.17 \citep{Los2}.
It might, however, help explain mismatches with the expected
theoretical growth rate that was found to be proportional to $\Gr$
\citep{KBKMR13} and that was determined from recent DNS \citep{Jab2}.

An important question is whether NEMPI will really be strong enough
to produce sunspots with super-equipartition strength.
It has always been clear that NEMPI can only work for a magnetic field
strength that is a small fraction of the local equipartition field value.
However, super-equipartition fields are produced if the magnetic
field is vertical \citep{BKR13}.
Subsequent work showed quantitatively that NEMPI does indeed work at
subequipartition field strengths, but since mass flows mainly along
magnetic field lines, the reduced pressure leads to suction which tends
to evacuate the upper parts of the tube \citep{Jab2}.
This is similar to the ``hydraulic effect'' envisaged by \cite{Par76},
who predicted such downflows along flux tubes.
In a later paper \cite{Par78}, gives more realistic estimates, but
the source of downward flows remained unclear.
Meanwhile, the flux emergence simulations of \cite{RC14} show at first
upflows in their magnetic spots (see their Fig.~5), but as the spots
mature, a downflow develops (see their Fig.~7).
In their case, because they have convection, those downflows can also be
ascribed to supergranular downflows, as was done by \cite{SN12}.
Nevertheless, in the isothermal simulations of \cite{BKR13,Jab2},
this explanation would not apply.
Thus, we now know that the required downflows {\em can} be caused by
NEMPI, but we do not know whether this is also what happens in the Sun.

Coming back to our paper, where NEMPI is coupled to a dynamo,
the recent work of \cite{MBKR14} is relevant
because it shows that intense bipolar spots can be generated in an
isothermal simulation with strongly stratified non-helically driven
turbulence in the upper part and a helical dynamo in the lower part.
The resulting surface structure resembles so-called $\delta$ spots
that have previously only been found in the presence of strongly
twisted and kink-unstable flux tubes \citep{LDFL98}.
While the detailed mechanism of this work is not yet understood, it
reminds us that it is too early to draw strong conclusions about NEMPI
as long as not all its aspects have been explored in sufficient detail.

\begin{acknowledgements}
This work was supported in part by the European Research Council under the
AstroDyn Research Project No.\ 227952,
the Swedish Research Council under the grants 621-2011-5076 and 2012-5797,
the Research Council of Norway under the FRINATEK grant 231444 (AB, IR),
and a grant from the Government of the Russian Federation under
contract No. 11.G34.31.0048 (NK, IR).
We acknowledge the allocation of computing resources provided by the
Swedish National Allocations Committee at the Center for
Parallel Computers at the Royal Institute of Technology in
Stockholm and the Nordic Supercomputer Center in Reykjavik.
\end{acknowledgements}


\end{document}